\documentclass[twocolumn, 12pt]{iopart}

\usepackage{iopams}  
\usepackage{graphicx}
\usepackage{epstopdf}
\usepackage{epsfig}
  
\begin{document}

\title[Temperature dependence of dielectric constants in TiN]{Temperature dependence of dielectric constants in Titanium Nitride}

\author{S.Tripura Sundari*, R.Ramaseshan, Feby Jose, S.Dash, A.K.Tyagi}

\address{Surface and Nanoscience Division, MSG, IGCAR, Kalpakkam, India 603102}
\ead{*sundari@igcar.gov.in}

\begin{abstract}
The temperature dependence of optical constants of titanium nitride thin film is investigated using spectroscopic ellipsometry between 1.4 to 5 eV in the temperature range 300 K to 650 K in steps of 50 K.  The real and imaginary parts of the dielectric functions $\varepsilon_1$(E) and $\varepsilon_2$(E) increase marginally with increase in temperature. A Drude Lorentz dielectric analysis based on free electron and oscillator model are carried out to describe the temperature behavior. The parameters of the Lorentz oscillator model also showed that the relaxation time decreased with temperature while the oscillator energies increased. This study shows that owing to marginal change in the refractive index with temperature, titanium nitride can be employed for surface plasmon sensor applications even in environments where rise in temperature is imminent. 

\end{abstract}

\maketitle

\section{Introduction}
Transition metal nitride thin films such as titanium nitride have been employed for a variety of hard, optical, anti-reflection, anti-static coating applications and as diffusion barriers. They are also employed as electrodes for semiconductor technology, owing to their exceptional combination of physical properties such as high hardness, elasticity, luster and metallurgical stability. It is widely used as a wear resistant coating on tools because of its mechanical resistance, low friction coefficient and high melting point. It is also used as a diffusion barrier in semiconductor metallization systems owing to its high chemical stability and low resistivity. Due to its gold like appearance, it is used as an alternative material to gold \cite{Beck}. The high Drude like reflectance in the infra red and high absorption in the visible and infra red near region makes it a good selective absorber.  
Fundamental studies have been carried out on stoichiometric and non-stoichiometric nanocrystalline thin films using X-ray analysis and spectroscopic ellipsometry with respect to identifying the electron scattering mechanisms. Spectroscopic ellipsometry was employed to investigate the optical, electronic and transport properties of TiN thin films synthesized by reactive magnetron sputtering\cite{Patsalas}. The amorphous form of TiN films exhibited high electrical resistivity, high transmission of light in the visible range and a lack of metallic brilliance\cite{Adam}.  In the thin films synthesized by rf magnetron sputtering, a shift in the plasma band from 4.83 to 2.47 eV was observed and this was attributed to the change in stoichiometry of the films \cite{Vasu}. In situ ellipsometric studies were carried out to understand the growth of ultra thin films synthesized by plasma assisted atomic layer deposition technique\cite{Langereis}. The evolution of the dielectric function with thickness was modeled based on Drude Lorentz oscillator parameterization. Apart from silver \cite{Sahin}, recently, TiN is also being explored as a material for plasmonic applications in the visible and near infra red wavelengths of the electromagnetic spectrum \cite{Naik}. It is known that the characteristics of SPR employing silver are affected by the temperature of the environment through temperature dependent refractive indices. It would therefore be interesting to explore the temperature dependence of dielectric functions of TiN thin films. Ellipsometry has been exploited as a unique research tool to determine the optical properties of materials in a non-invasive and non-contact manner with a high degree of accuracy. The real and imaginary parts of the complex dielectric function of a material can be determined directly without the inversion of Kramer’s Kronig relation using this technique.  In the current article, we report on the evolution of dielectric function of TiN for temperatures ranging from 300 K to 650 K in steps of 50 K using a spectroscopic ellipsometer in the  energy range 1.4 to 5 eV. 

\section{Experimental}
Titanium Nitride thin films were synthesized by RF reactive magnetron sputtering technique, using MECA 2000 (France) sputtering system. For deposition, Si $<100>$ substrates were initially loaded into the coating chamber which was pumped down to a base pressure of 1 x $10^{-6}$ mbar. Prior to deposition, the substrates were subjected to in-situ $Ar^+$ ion bombardment in order to ensure good adhesion of the film to the substrate. The targets were also sputter cleaned before deposition so as to render it clean and expose fresh surface for sputtering. Thereafter, a thin layer of Ti (~50 nm) nm was deposited as a buffer layer so as to improve the adhesion between the substrate and to- be-deposited TiN thin film. A 50 mm diameter, 4N pure Ti disk was sputtered at an operating RF power of 300W. The ratio of sputtering gas (Ar : 5N pure) and reactive gas ($N_2$ : 5N pure) was kept constant at 4:1 for all coatings. The target to substrate distance was maintained at 100 mm while the dynamic pressure of  1 x $10^{-2}$  mbar was maintained during the deposition process. The films were deposited at a rate of 7nm/min. The ellipsometric parameters of the films were measured by a SOPRA ESVG rotating polarizer type spectroscopic ellipsometer in the energy range 1.4 to 5.0 eV at an angle of incidence of $75\,^{\circ}{\rm C}$.  For the high temperature ellipsometric measurements, the samples were placed in a resistive heating stage in a high vacuum chamber which is capable of being maintained at a vacuum of $\sim 10^{-7}$ mbar. The pumping was carried out by oil free pumps and the accuracy of temperature was maintained at $\sim \pm 0.1\,^{\circ}{\rm C}$. The experimental procedures mandatory for the correction due to windows etc were carried out using a Si standard crystal. It is to be noted that the line-shape and spectral positions of the $E_1$ and $E_2$ transitions are strongly dependent on temperature and this has been used for the above purpose.

\section{Results and Discussion}

\subsection{Analysis of dielectric function }
Ellipsometry is a non invasive, non-contact technique that has been used to investigate the optical properties of materials. It is based on the principle of detection of change in polarization of incident light after reflection from the system under investigation. It measures $\rho$, the ratio between the parallel and perpendicular components of the complex reflection coefficients through which the dielectric functions of the material under study is investigated. The complex dielectric function $\varepsilon$ is directly obtained from the measured ellipsometric parameters using the relations 

\begin{equation}
\varepsilon(E) = N^{2}_o\left[sin^{2}\phi+\left[\frac{1-\rho}{1+\rho}\right]^{2}sin^{2}\phi \ tan^{2}\phi\right]
\end{equation} 

where $N_o$ is the refractive index of the ambient, E is the energy of the incident electromagnetic radiation and $\phi$ is the angle of incidence.  The imaginary part of the dielectric function is directly related to the conduction electron density for metals and to the density of states for the interband transitions. The optical properties of metals can be described in terms of intraband and interband transitions depending on the frequency $\omega$ (E=h$\nu$) of the incident electromagnetic waves. The contribution from the intraband and interband to the dielectric function is summed by the Drude-Lorentz theory whose expression is given below
  
\begin{equation}
\varepsilon = \varepsilon_\infty - \left(\frac{\omega^2_{pu}}{\omega^2 - i\Gamma_{D}\omega}\right)+\sum_{j=1}^2\left(\frac{f_j\omega^2_{oj}}{\omega^2_{oj}-\omega^2+i\gamma_j\omega}\right)
\end{equation} 
 
In the above equation, $\varepsilon_\infty$ is a background constant which is larger than unity that takes into account the contribution from higher energy transitions. In the lower spectral region where $\omega\tau \gg$1, the Drude model can be approximated and the real and imaginary parts of the dielectric function can be expressed as

\begin{equation}
\varepsilon_1 \cong \varepsilon_\infty - \frac{\omega^2_{pu}}{\omega^2} = \frac{\lambda^2}{\lambda^2_{pu}}
\end{equation}

and         
\begin{equation}
\varepsilon_2 \cong \frac{\omega^2_{pu}}{\omega^3\tau}\cong \frac{\lambda^3}{\lambda^2_{pu}\lambda_\tau}=\frac{\varepsilon_\infty-\varepsilon_1}{\omega\tau}
\end{equation}

where $\omega_{pu},\ \omega,\ \lambda,\ \varepsilon_\infty,\ \lambda_{pu},\ \lambda_{\tau}\ and\ \tau$ are the unscreened plasma frequency, frequency, wavelength of incident light, core polarizibility, unscreened plasma wavelength, relaxation wavelength  and relaxation time of conduction electrons, respectively.  The drude term is characterized by an unscreened plasma frequency $\omega_{pu}$ and the damping factor $\Gamma_{D}$ . The damping factor is the contribution due to the scattering of electrons and according to free electron theory, it is the inverse of the electron relaxation time $\tau$ given by $\tau  =\hbar /\Gamma_{D} $.  As per the approximation of the Drude model (3), $-\epsilon_1$ should be a linear function of $\lambda^2$ from which the inverse of the square root of the slope and intercept yield unscreened plasma frequency and core polarizibility, respectively.  

The Lorentz oscillators describe the interband transitions and provide insight into the variation of the oscillator strength $f_j$, energies $\omega_{oj}$ and broadening of the $i^{th}$ oscillator, respectively. The Lorentz oscillator fitting is not unique since it results in a relatively large uncertainty in the parameters whereby, comparison between results is difficult. However, it has been employed in many investigations in order to understand the optical trends as a function of some physical parameter. In the present studies, we examine the parameters of the oscillators as a function of temperature. 

\subsection{Behavior of dielectric function at room temperature}

\begin{figure}[h]
\begin{center}
\includegraphics[width=0.55\textwidth]{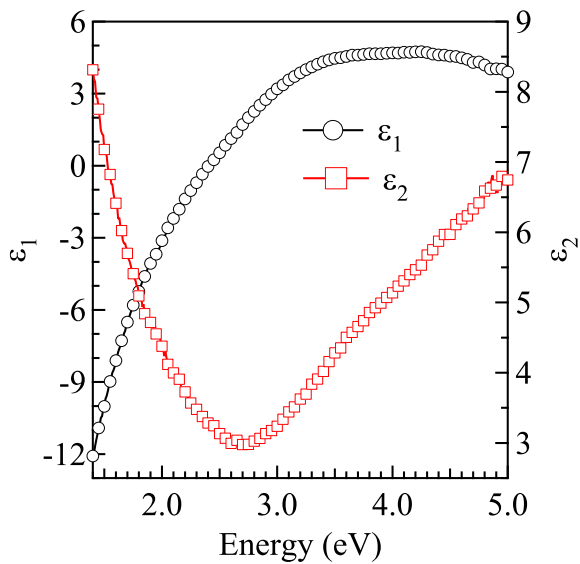}
\caption{\label{Fig1}(Color online) Real and imaginary parts of the dielectric function measured at 300 K.    }
\end{center}
\end{figure}

Figure 1 shows the real $\epsilon_1$ and imaginary $\epsilon_2$ parts of the dielectric function of TiN film measured at room temperature. The spectrum is dominated by the highly polarizable free charge carriers, leading to the sharply increasing $\epsilon_1$ and decreasing $\epsilon_2$ with increasing photon energy. It is observed that the magnitude of $ <\varepsilon_1 (E)> $ shows a typical free electron like behavior, increasing from a large negative value of -12.66 at 1.4 eV to a maximum value of ~ 4.7 near 4.2 eV. The $ <\varepsilon_2 (E)>$  decreases with increasing energy from a large value (8.7 at 1.4 eV) to a minimum (2.97 at 2.68 eV) at an energy somewhat larger than the screened plasmon mode. This indicates that below this energy, the majority contribution is from intraband transitions due to free electrons. The $\epsilon_2$ spectra are nearly comparable to those reported in the literature \cite{Patsalas, Ji, Schlegel, Adachi}. indicating high density of the films. It is to be pointed out that, higher the value of $\epsilon_1$, the higher the density of films as inferred from density calculations based on XRR \cite{Patsalas}.  The intersection of $\epsilon_1$ with a positive slope near 2.42 eV is indicative of a longitudinal excitation mode i.e., screened plasmon response. This intersection at a value of $\epsilon_1$= 0 coincides with a maximum of the calculated energy loss function.  The collective behavior of electrons can be judged from the energy loss spectrum, where $\epsilon_2$ drops to a minimum while the $\epsilon_1$ crosses through zero.  Ideally, in the region of domination by interband transitions, the energy loss function should exhibit a spike like behavior. However, the loss function is usually broad due to background and damping of single particle transitions and is of the order of few eV. The equivalence between the transverse and longitudinal dielectric constant in the energy range of 1.4 to 5 eV enables one to compare between loss spectrum calculated from the optical spectra obtained by ellipsometry and those from energy loss spectrum of fast electrons travelling in a crystalline material. The energy loss function is evaluated from the expression 

\begin{equation}
-Im(1/\varepsilon) = \frac{\varepsilon_2} {\varepsilon^2_1 + \varepsilon^2_2}
\end{equation}

\begin{figure}[h]
\begin{center}
\includegraphics[width=0.55\textwidth]{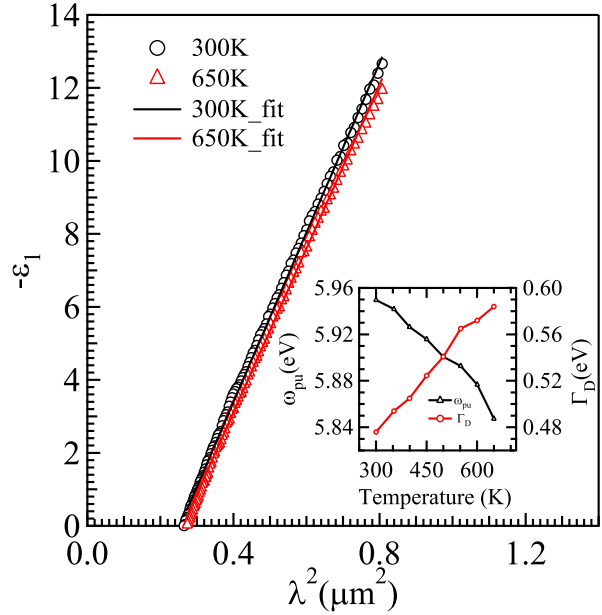}
\caption{\label{Fig2}(Color online) A plot of $-\varepsilon \ Vs \ \lambda^2$ shown for two representative measurement temperatures namely 300 K and 650 K  (Inset: variation of unscreened plasma frequency ($\omega_{pu}$) and Drude broadening ($\Gamma_D$) with temperature T).}
\end{center}
\end{figure}

and involves both one electron excitations and collective excitations. The energy loss spectra at room temperature shows a peak near 2.5 eV, which is close to the strong plasmon resonance. Our results are in excellent agreement with the golden colored specimen of ref. \cite{Adachi}. It is to be pointed out that the energy (or screened plasma frequency) of longitudinal excitation mode is known to depend on the stoichiometry of Ti and N as has been pointed out by Ji Hoon \emph{et al.} \cite{Ji}. Our results also compare well with TiN films synthesized by reactive sputter techniques\cite{Adachi}. The contribution to the optical response in TiN arises from the electronic band structure in which Fermi level lies within the d-band and accounts for the free electrons at the d-band at energies below 2.5 eV.  Figure 2 shows a plot of $-\varepsilon_1 \ Vs \ \lambda^2$ for the measurement carried out at 300 K.  The unscreened plasma frequency inferred from the plot is 5.39 eV. Based on the electronic band structure \cite{Ern}, the interband transitions in TiN take place at $\Gamma_{15}  \rightarrow   \Gamma_{12}$ , $X_5 \rightarrow X_2$ and  $L_3 \rightarrow L_{3'}$ as well as K and W symmetry lines of the Brillouin zone points of the Brillouin zone corresponding to absorptions ~2.3, 3.9 and 5.6 eV,  respectively.  The lowest band gap transition ($\Gamma_{25'}  \rightarrow   \Gamma_{12}$)   is reported to be very weak and is located close to 1.0 eV and this is the transition that is responsible for the gold like color of TiN thin films. The other two transitions are interpreted as due to the transitions between hybridized N p-Ti d bands below the Fermi level $E_F$ and Ti s-d bands above $E_F$. 

\begin{figure}[h]
\begin{center}
\includegraphics[width=0.55\textwidth]{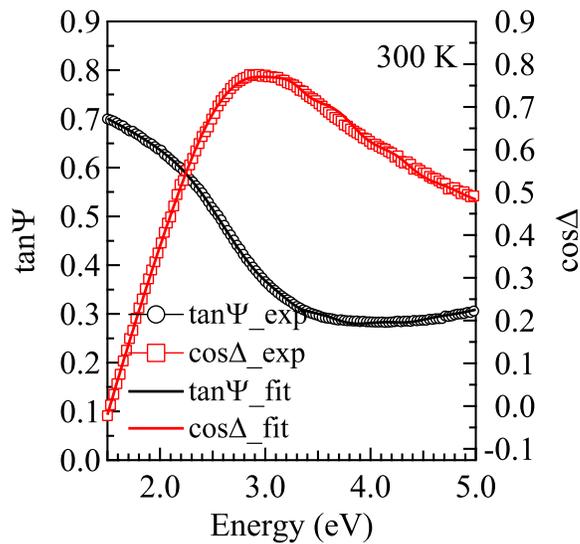}
\caption{\label{Fig3}(Color online) Experimentally measured ellipsometric parameters for the room temperature measurement along with the best fit.}
\end{center}
\end{figure}

A linear regression analysis employing the Levenberg Marquadt algorithm was carried out on the measured ellipsometric data in which the parameters of the Drude-Lorentz model are varied until the mean square deviation $\chi^2$ (defined below) between computed and experimental values was below $\sim 10^{-4}$. The mean square deviation ($\chi^2$) is defined as

\begin{equation}
\chi^2 = \left[\frac{1}{P-M-1}\right] \sum_{j=1}^2 \left[ \left(\tan\psi_{j, expt} - \tan\psi_{j, comp}\right)^2  + \left(\cos\Delta_{j, expt} - 
\cos\Delta_{j, comp}\right)^2\right] 
\end{equation}

\begin{figure}[h]
\begin{center}
\includegraphics[width=0.55\textwidth]{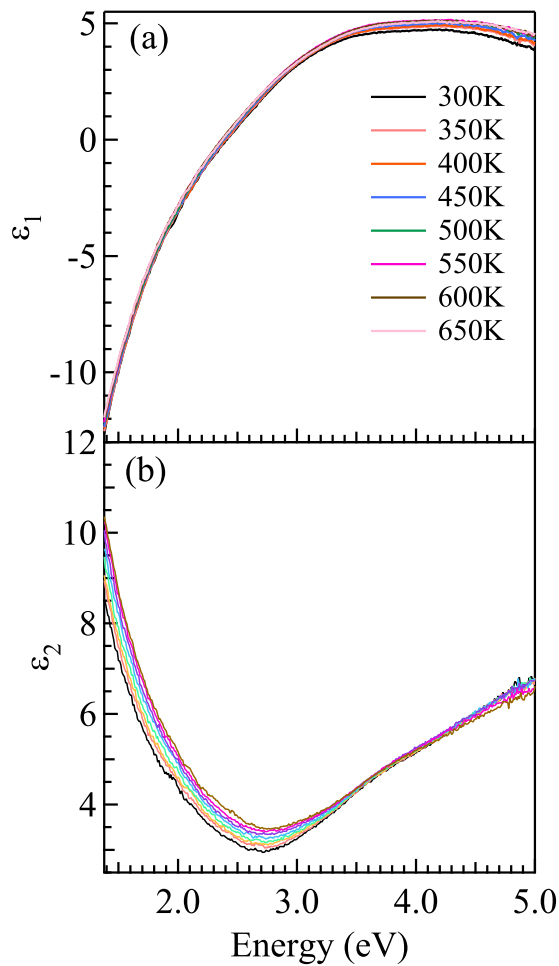}
\caption{\label{Fig4}(Color online) Plot of the real and imaginary parts of the dielectric constants for measurements temperatures 300 K to 650 K}
\end{center}
\end{figure}

where P is the number of data points and M is the number of variables to be fitted, $\psi_{expt}$  and $ \psi_{comp}$ are experimental  and computed values of $\psi$. In the present experiment, the variation of the dielectric functions is described by the parameters of the Drude-Lorentz oscillator model, wherein a three layer model is employed (c-Si/TiN/ambient). The reference refractive index for c-Si was taken from ref. \cite{Aspnes}, while the TiN layer was modeled based on by Drude Lorentz model. In the current analysis, we have used two oscillators located near 3 eV and 6 eV for the input parameters. The best fit for the energies of the two oscillators were found to be at 3.14 eV and 6.02 eV for the data measured at room temperature and is shown in Figure 3. The thickness of the films as evaluated from ellipsometry was 80 nm. 

\begin{figure}[h]
\begin{center}
\includegraphics[width=0.55\textwidth]{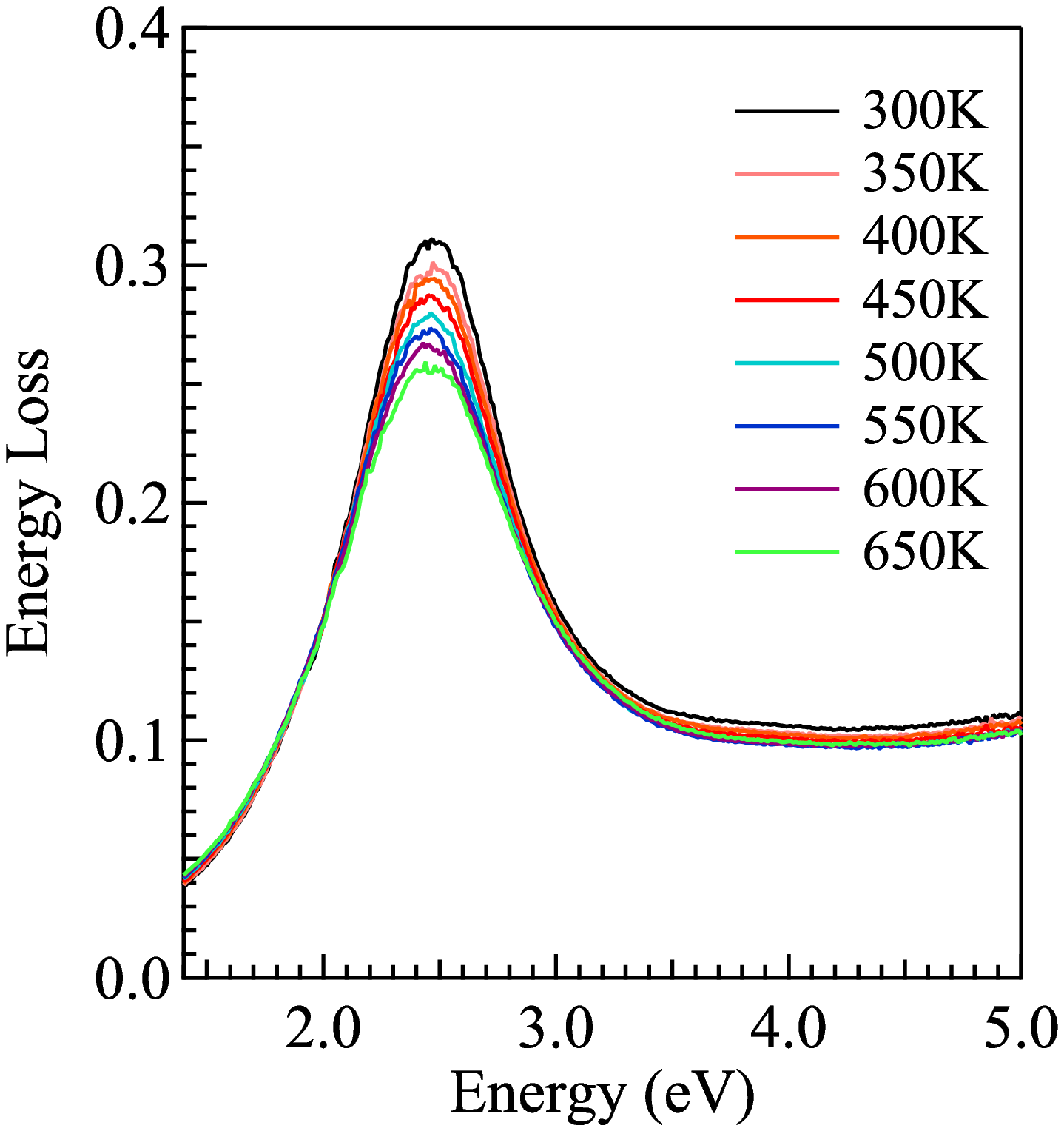}
\caption{\label{Fig5}(Color online) Energy loss function for different measurement temperatures}
\end{center}
\end{figure}

\subsection{Behavior of dielectric functions with temperature}
Figure 4(a) and 4(b) show the real and imaginary parts of the complex dielectric function obtained from the ellipsometric parameters measured between 300 K to 650 K is steps of 50 K.  It is seen from the figures that both $\varepsilon_1(E)$   and $\varepsilon_2(E)$   increases marginally with increase in temperature. The increase in the magnitude of $\varepsilon_1(E)$ is a consequence of the decrease in absorption thereby leading to increase in propagation. The increase in $\varepsilon_2(E)$ in the red and infra red region energy range is due to enhanced damping of conduction electrons due to the increased number of thermal phonons. In general, with reference to metals, three mechanisms, which are inter-related to each other have been used to explain the temperature dependence of the dielectric functions. These are isotropic volume thermal expansion, increased electron-phonon population, Fermi-distribution broadening and shift of the Fermi level. The contribution of all these factors for a particular photon energy depends on the variation of electronic structure with temperature and the transition matrix amplitudes. The thermal expansion in particular, causes modifications of electron energy band through the smearing of electron density. Therefore, it is of interest to follow the temperature dependence of unscreened plasma frequency, which is a direct reflection of the density of electrons. Based on eq. 3, we have plotted $-\varepsilon_1 \ Vs \ \lambda^2$ as shown in Figure 2, for two representative temperatures 300 K and 650 K in order to extract the unscreened plasma frequency. It is indeed linear and the unscreened plasma frequency shows a marginal decrease from 5.95 eV to 5.85 eV. The behavior of $\omega_{pu}$ has been shown to be sensitive to composition, temperature and sample synthesis conditions.  The $\omega_{pu}$ is affected by the conduction electron density which in turn depends on the mass density through the relation $\omega_{pu}$ = $ne^2/m^*\varepsilon_0$, where $m^*$ is the conduction electron effective mass and $\varepsilon_0$ is the permittivity of free space. In the study of TiN nanocrystalline thin films, an increase in $\omega_{pu}$ was observed with increase in bias voltage and interpreted as an enhancement of the metallic character in terms of the electrical conductivity\cite{Patsalas}.  The increase in $\omega_{pu}$ in ref.\cite{Patsalas} is attributed to increase in mass density and grain growth with increase in substrate temperature. While $\omega_{pu}$ shows an increase as in the case referred above, a decrease in $\omega_{pu}$ has also been observed.  In the investigation of temperature dependence of Ag (100)\cite{Rocca},  the decrease in $\omega_{pu}$ has been attributed to thermal expansion coefficient.  In the study of cubic $TiN_x$, a decrease in $\omega_{pu}$ was observed with increase in nitride composition \cite{Ji} and was interpreted as a decrease in conduction electron density which is mostly contributed by Ti atoms. Therefore, in this case, the decrease of $\omega_{pu}$ with increase in nitride composition is construed as strengthening of covalent bonding thereby leading to a reduction in the conduction electron density. The shift in volume plasmon resonances have been reported to be 105 meV and 120 meV for Al and Pb respectively for temperature variation from 10K to 295 K.  A strong dependence of $\omega_{pu}$ on the sample conditions was also reported by Parmigiani $\emph et al$ \cite{Parmigiani}. The decrease in $\omega_{pu}$ with increasing substrate temperature has been observed in the case of Cu and Ag thin films.  In the present experiment, a decrease in $\omega_{pu}$ is attributed to thermal expansion in lattice thereby, implying a decrease in the conduction electron density.

The broadening parameter $\Gamma_D(=\hbar/\tau)$ which is the inverse of relaxation time depends on the phonon contribution and microstructural parameters such as  static impurities, defect density, grain boundary, grain sizes etc \cite{Savaloni}. In the present experiment, it is observed that $\Gamma_D$ increases with increase in temperature (inset of Figure 2) implying a decrease in electron relaxation time $\tau$. This decrease in $\tau$ is due to the phonon contribution owing to rise in temperature. Similar effects have been found in the case of temperature dependence of Cu and Ni by Johnson et al\cite{Johnson}. The energy loss function (ELF) calculated from eq. 5, (Figure 5) shows that there is a significant decrease in peak height as compared to peak position and width. The peak height of the ELF is $1/\varepsilon_2(E_0)$ and the magnitude of the full width at half maximum is $2\varepsilon_2/\left[d\varepsilon_1(E_0)/dE\right]$ at $E_0$ . Therefore, while the height of ELF is determined only by the magnitude of $\varepsilon_2$, the width is decided by both $\varepsilon_2$ and slope of $\varepsilon_1$ that varies with energy. The marginal increase in damping in ELF with increase in T is a consequence of the marginal increase in magnitude of $\varepsilon_2$ in the region where $\varepsilon_1 \sim 0$. It is seen that the changes in $\varepsilon_2$ are more significant in comparison to that of $\varepsilon_1$ thereby resulting in more prominent changes to peak height rather than to peak width. 

\begin{figure}[h]
\begin{center}
\includegraphics[width=0.55\textwidth]{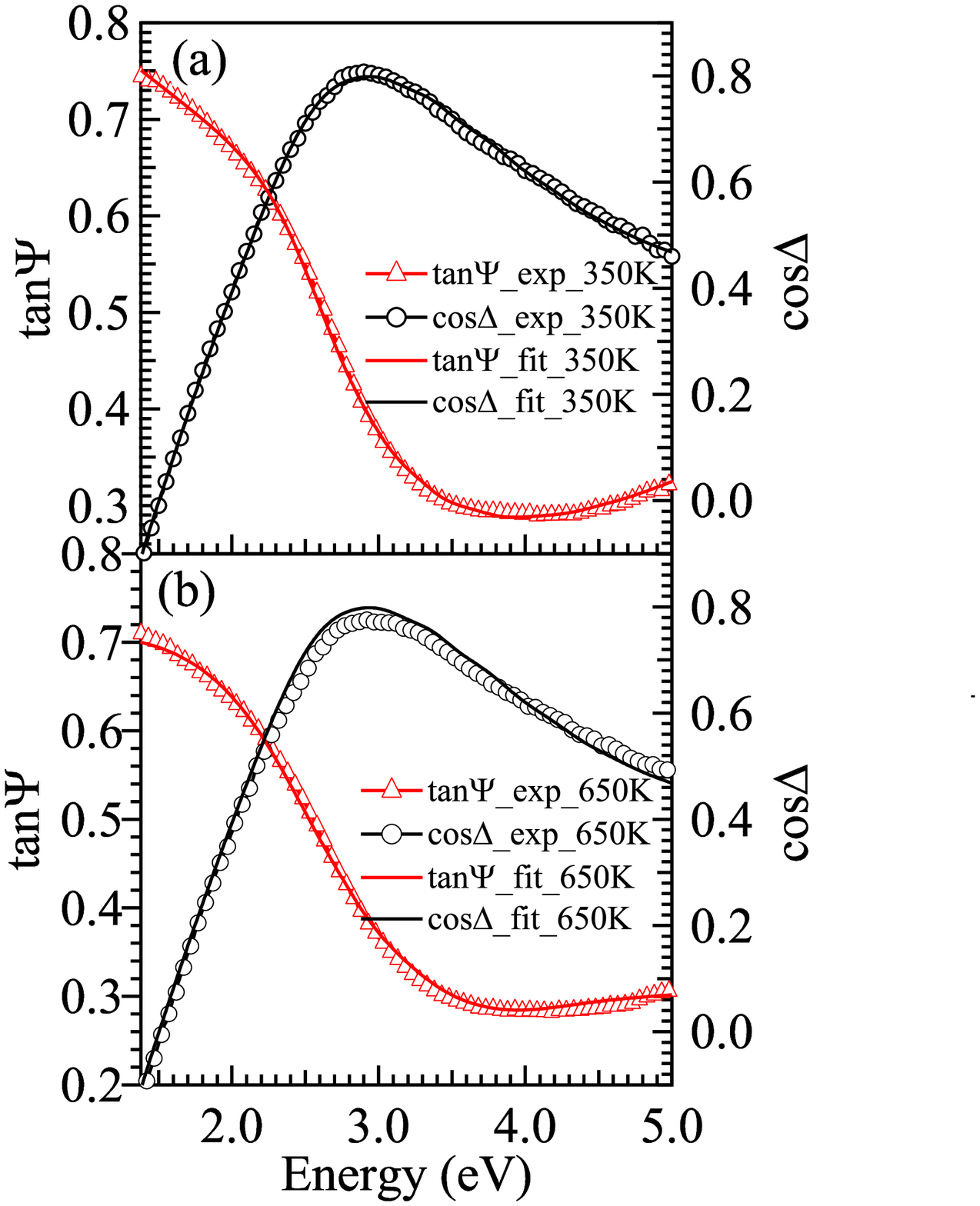}
\caption{\label{Fig6}(Color online)A plot of experimental ellipsometric parameters and the corresponding fit for two representative specimens (350K and 650K) }
\end{center}
\end{figure}

\begin{figure}[h]
\begin{center}
\includegraphics[width=0.55\textwidth]{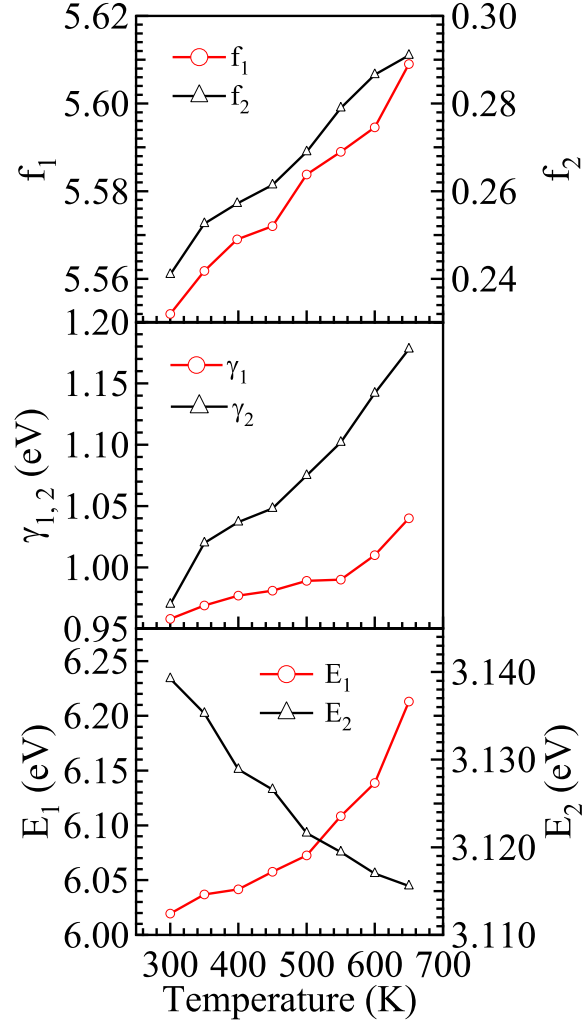}
\caption{\label{Fig7}(Color online)Variation of parameters of Lorentzian oscillators  (viz. energy, amplitude and broadening) with temperature

}
\end{center}
\end{figure}

The parameters of the Lorentz oscillators were extracted by fitting the experimental data to eq. 2 using the condition of eq. 6. A representative fit is shown in Figure 6(a) and 6(b) for measurements carried out at 350 K and 600 K, respectively. It is seen from the figure that the fits are excellent. As far as the variation of parameters of Lorentz oscillators are concerned (Figure 7), increase in temperature shows a marginal shift of oscillator $E_{01}$ from 3.14 eV to 3.11 eV, while the second oscillator at $E_{02}$ shifts to higher energy from 6.01 to 6.2 eV. The position of the oscillators in general, agrees well with those already reported in except for small variations. They are found to vary depending on the microstructure of the system under investigation. In the case of TiN thin films synthesized by plasma assisted atomic layer deposition, oscillators were located at 3.8 and 5.6 eV, respectively \cite{Langereis}, while for nanocrystalline thin films, the oscillators were best fit between $3.6 - 3.72$ and $5.2 - 6.3$ eV depending on the bias voltage.  The difference in the positions of the oscillators could be due to the differences in the microstructure \cite{Patsalas}. The amplitudes shown in ($f_1$ from 5.55 (300 K) to 5.60 (650 K) and $f_2$ : 0.24 (300K) to 0.29 (650K)) and broadening  ($\gamma_1$ from 0.97 eV (300 K) to 1.18 eV (650 K) and $\gamma_2$ : 0.96 eV (300K) to 01.04 eV (650K) of  both the oscillators show a marginal increase with temperature. The increase in broadening and oscillator strengths of the oscillators are indicative of increase in the strength of $X_5 \rightarrow X_2$ and $L_3 \rightarrow L_{3'}$ transitions and possibly other transitions that take place along the K and W directions\cite{Bohm}. Similar changes in oscillator strengths and amplitudes have been observed in ref.\cite{Patsalas, Langereis}.

\subsection{Conclusion}
The high temperature optical properties of titanium nitride thin films are presented for photon energies ranging from 1.4 to 5 eV between 300 K and 650 K in steps of 50 K using spectroscopic ellipsometry. The real and imaginary parts of the dielectric function showed marginal increase with temperature.  The ellipsometric data are modeled based on Drude Lorentz dispersion law. The unscreened plasma frequency decreased with increase in temperature, while the relaxation time of electrons showed a marginal decrease. Similarly, the parameters of the oscillators viz., energy, strength and broadening also showed marginal changes. The energy loss spectrum -Im (1/$\varepsilon$) showed a significant decrease in the peak height, as compared to its peak position and peak width. The marginal and small change in the optical properties with temperature shows that titanium nitride can be employed in high temperature environments as surface plasmon resonance sensors 

\subsection{Acknowledgments}
The authors would like to thank Kunuku Srinivas for his support in experimental work.

\end{document}